\documentclass[aps,floats]{revtex4}
\usepackage{amsmath,amssymb}
\usepackage{graphicx,epsfig}

\begin{document}
\bibliographystyle {plain}

\def\oppropto{\mathop{\propto}} 
\def\opmin{\mathop{\min}} 
\def\opmax{\mathop{\max}} 
\def\opsimeq{\mathop{\simeq}}
\def\opoverderline{\mathop{\overline}}
\def\operarrow{\mathop{\longrightarrow}}
\def\opsim{\mathop{\sim}}

\def\fig#1#2{\includegraphics[height=#1]{#2}}
\def\figx#1#2{\includegraphics[width=#1]{#2}}


\title{ Anderson transitions : multifractal or non-multifractal statistics  \\
of the transmission as a function of the scattering geometry  } 


 \author{ C\'ecile Monthus and Thomas Garel }
\affiliation{Institut de Physique Th\'{e}orique, CNRS and CEA Saclay
 91191 Gif-sur-Yvette cedex, France}

\begin{abstract}
The scaling theory of Anderson localization is based on a global conductance $g_L$ that remains a random variable of order $O(1)$ at criticality. One realization of such a conductance is the Landauer transmission for many transverse channels. On the other hand, the statistics of the one-channel Landauer transmission between two local probes is described by a multifractal spectrum that can be related to the singularity spectrum of individual eigenstates. To better understand the relations between these two types of results, we consider various scattering geometries that interpolate between these two cases and analyse the statistics of the corresponding transmissions. We present detailed numerical results for the power-law random banded matrices (PRBM model). Our conclusions are : (i) in the presence of one isolated incoming wire and many outgoing wires, the transmission has the same multifractal statistics as the local density of states of the site where the incoming wire arrives; (ii) in the presence of backward scattering channels with respect to the case (i), the statistics of the transmission is not multifractal anymore, but becomes monofractal. Finally, we also describe how these scattering geometries influence the statistics of the transmission off criticality.

\end{abstract}

\maketitle

\section{ Introduction }

Within the field of Anderson localization \cite{anderson}
 (see the reviews \cite{thouless,souillard,bookpastur,
Kramer,janssenrevue,markos,mirlinrevue}), one main achievement
has been the formulation of a scaling theory \cite{scaling}
describing the renormalization flow of the typical global conductance
$g_L$ as a function of the linear size $L$ of the system. 
The critical point then corresponds to a finite conductance,
whereas the delocalized phase corresponds to the growing RG flow
$g_L \sim L^{d-2}$ in dimension $d>2$, and the localized phase corresponds
to the decaying RG flow $g_L \sim e^{- (cst)L}$.
Although this scaling theory was first formulated in terms of 
the Thouless definition of conductance, based on the sensitivity
to boundary conditions, it was then realized that a more 
straightforward definition of a global conductance 
can be obtained via the quantum scattering theory
where the quantum transmission $T_L^{m.c.}$ is given
by the many-channel (m.c.) Landauer formula
\cite{newscaling,fisherlee,buttiker,stone}
that generalizes the one-dimensional case \cite{landauer,
anderson_lee,luck}.

This scattering definition has the advantage  
to have a well defined meaning for each disordered sample,
and thus one may study its probability distribution
over the samples \cite{newscaling} (see \cite{markos} for a review of the
results concerning this distribution both in the localized phase
and at criticality). 
From the point of view of the scaling theory, the important point
is that at criticality, the many-channel Landauer transmission
 $T_L^{m.c.}$ remains a random variable of order $O(1)$ as $L \to +\infty$. 
On the other hand, it is now well established that 
 individual critical eigenfunctions are multifractal 
(see the reviews \cite{janssenrevue,mirlinrevue}).
As a consequence, the one-channel
 Landauer transmission $T^{(1,1)}_L$ 
between two local probes is expected to display also
 multifractal properties at criticality \cite{janssen99,evers08,
us_twopoints}.
To better understand the relation between this multifractal statistics
of the one-channel Landauer transmission $T_L^{(1,1)}$
and the many-channel Landauer transmission $T_L^{m.c.}$,
we consider in this paper various scattering geometries
 that interpolate between the two in order to characterize the statistics
of the transmission in these intermediate conditions.
We present detailed numerical results for the Anderson critical points of
the Power-law Random Banded Matrix (PRBM) model, where the 
criticality condition is exactly known.

The paper is organized as follows. In section \ref{secmodel}, we 
recall the definition and some properties of the PRBM model. 
We then consider the statistics of the transmission at criticality
for the following scattering geometries :

- one incoming wire and one outgoing wire (section \ref{twopoint})

- one incoming wire and many outgoing wires (section \ref{1tomany})

- one incoming wire with many backward channels and many outgoing wires
(section \ref{backwards})

For completeness in section \ref{seclocdeloc}, we describe 
how these scattering geometries influence the statistics of the transmission
off criticality.
 Our conclusions are summarized in section \ref{secconclusion}.
A brief reminder on multifractality of critical eigenfunctions
is given in Appendix A.

\section{ Reminder on the Power-law Random Banded Matrix (PRBM) model  }

\label{secmodel}

Beside the usual short-range Anderson tight-binding model
in finite dimension $d$, other models displaying Anderson localization
have been studied,
in particular the Power-law Random Banded Matrix (PRBM) model,
which can be viewed as a one-dimensional model with long-ranged
random hopping decaying as a power-law $(b/r)^a$ of the distance $r$
with exponent $a$ and parameter $b$.
Many properties of the Anderson transition at $a=1$ between localized 
states with integrable power-law tails($a>1$)
 and extended states ($a<1$) have been studied, in particular
the statistics of eigenvalues \cite{varga00,kra06,garcia06}, 
and the multifractality of eigenfunctions 
\cite{mirlin_evers,cuevas01,cuevas01bis,varga02,cuevas03,mirlin06},
including boundary multifractality \cite{mildenberger}.
The critical properties at $a=1$ depend continuously
on the parameter $b$, which plays a role analog
to the dimension $d$ in short-range Anderson transitions 
\cite{mirlin_evers} : the limit $b \gg 1$ corresponds to
weak multifractality ( analogous to the case $d=2+\epsilon$)
and can be studied via the mapping onto a non-linear sigma-model
 \cite{mirlin96},
whereas the case $b \ll 1$ corresponds to strong multifractality
( analogous to the case of high dimension $d$)
and can be studied via Levitov renormalization \cite{levitov,mirlin_evers}.
Other values of $b$ have been studied numerically  
\cite{mirlin_evers,cuevas01,cuevas01bis,varga02}.
The statistics of scattering phases, Wigner delay times
and resonance widths in the presence of one external wire 
have been discussed in \cite{mendez05,mendez06}.
Related studies concern dynamical aspects \cite{limadyn},
the case with no on-site energies \cite{lima},
and the case of power-law hopping terms in dimension $d>1$ \cite{potempa,
cuevas04,cuevas05}.

In this paper, we consider the PRBM in dimension $d=1$ 
either in the line geometry with two boundaries
or in the ring geometry with periodic boundary conditions,
in the presence of various
external wires to measure the transmission properties.

\subsection{ PRBM with boundaries (line geometry) }

\label{defline}

In the line geometry, the distance $r_{i,j}$ 
between two sites $(i,j)$ is simply
\begin{eqnarray}
r_{i,j} = \vert i-j \vert
\label{rijline}
\end{eqnarray}
The ensemble of power-law random banded matrices of size 
$L \times L$ is then defined as follows : 
  the matrix elements $H_{i,j}$ are independent Gaussian
variables of zero-mean $\overline{H_{i,j}}=0$ and of variance
\begin{eqnarray}
\overline{ H_{i,j}^2 } = \frac{1}{1+ \left( \frac{r_{i,j}}{b}\right)^{2a}}
\label{defab}
\end{eqnarray}

\subsection{ PRBM with periodic boundary conditions (ring geometry) }

\label{defring}

In the ring geometry with periodic boundary conditions,
 the appropriate distance $r_{i,j}$
between the sites $i$ and $j$ on a ring of $L$ sites 
is defined as \cite{mirlin_evers}
\begin{eqnarray}
r_{i,j}^{(L)} = \frac{L}{\pi} \sin \left( \frac{ \pi (i-j) }{L} \right)
\label{rijcyclic}
\end{eqnarray}
The  matrix elements $H_{i,j}$ are then defined as before
in terms of this distance (Eq. \ref{defab}).
The main property of this ring geometry is that all sites 
are equivalent, whereas in the line geometry there are
two boundaries at $i=1$ and at $i=L$.

\subsection{ Scattering transmission in the presence of external wires }

In the following, we will consider various scattering 
geometries, where the disordered sample
 is linked to one incoming wire, parametrized
by the half-line ($x \leq x^{in}$), and to
one or many outgoing wires,
parametrized
by the half-lines ($x_j \geq x^{out}_j$).
In each case, we are interested 
in the eigenstate $\vert \psi >$ that satisfies 
the  Schr\"odinger equation
\begin{eqnarray}
H \vert \psi > = E \vert \psi > 
\label{schodinger}
\end{eqnarray}
inside the disorder sample
 characterized by the random $H_{i,j}$,
and in the perfect wires
characterized by no on-site energy
 and by hopping unity between nearest neighbors.
Within these perfect wires, one requires the plane-wave forms
\begin{eqnarray}
\psi_{in}(x \leq x^{in}) && = e^{ik (x-x^{in})} +r e^{- i k (x-x^{in})} \nonumber \\
 \psi_{out}(x_j \geq x^{out}_j) && = t_j e^{ik (x_j-x^{out}_j)} 
\label{psiwires}
\end{eqnarray}
These boundary conditions define
 the reflection amplitude $r$ of the incoming wire
and the transmission amplitudes $(t_j)$ of the outgoing wires.

The Landauer transmission $T$ of a subset $J$
of the outgoing wires defined as 
\begin{eqnarray}
T(J)\equiv  \sum_{j \in J} \vert t_j \vert^2 
\label{deftotaltrans}
\end{eqnarray}
is then a number in the interval $[0,1]$.
The various scattering geometries considered below
differ in the number of the possible out-going wires
and in the subset $J$ used to measure the transmission.

To satisfy the Schr\"odinger 
Eq. \ref{schodinger} within the wires described by
 Eq. \ref{psiwires}, one has the following relation between
the energy $E$ and the wave vector $k$  
\begin{eqnarray}
 E=2 \cos k  
\label{relationEk}
\end{eqnarray}
To simplify the discussion, we will focus in this paper on the case of
zero-energy $E=0$ (wave-vector $k=\pi/2$) that corresponds to the center of the band.

In the following, we study numerically
 the statistical properties
of the Landauer transmission $T$ for systems of sizes $50 \leq L \leq 1800$
with corresponding statistics of $10.10^8 \geq n_s(L) \geq 2400$
 independent samples.
For typical values, the number $n_s(L)$ of samples is sufficient
even for the bigger sizes, whereas for the measure of multifractal
spectrum, we have used only the smaller sizes where the statistics
of samples was sufficient to measure correctly the rare events.
To measure numerically multifractal spectra,
we have used the standard method based on
 $q$-measures of Ref. \cite{Chh}
(see Appendix B of \cite{us_twopoints} for more details).

\section{ Statistics of the one-channel transmission 
$T_L^{(1,1)}$ }

\label{twopoint}

\subsection{ Scattering geometry }

\begin{figure}[htbp]
 \includegraphics[height=12cm]{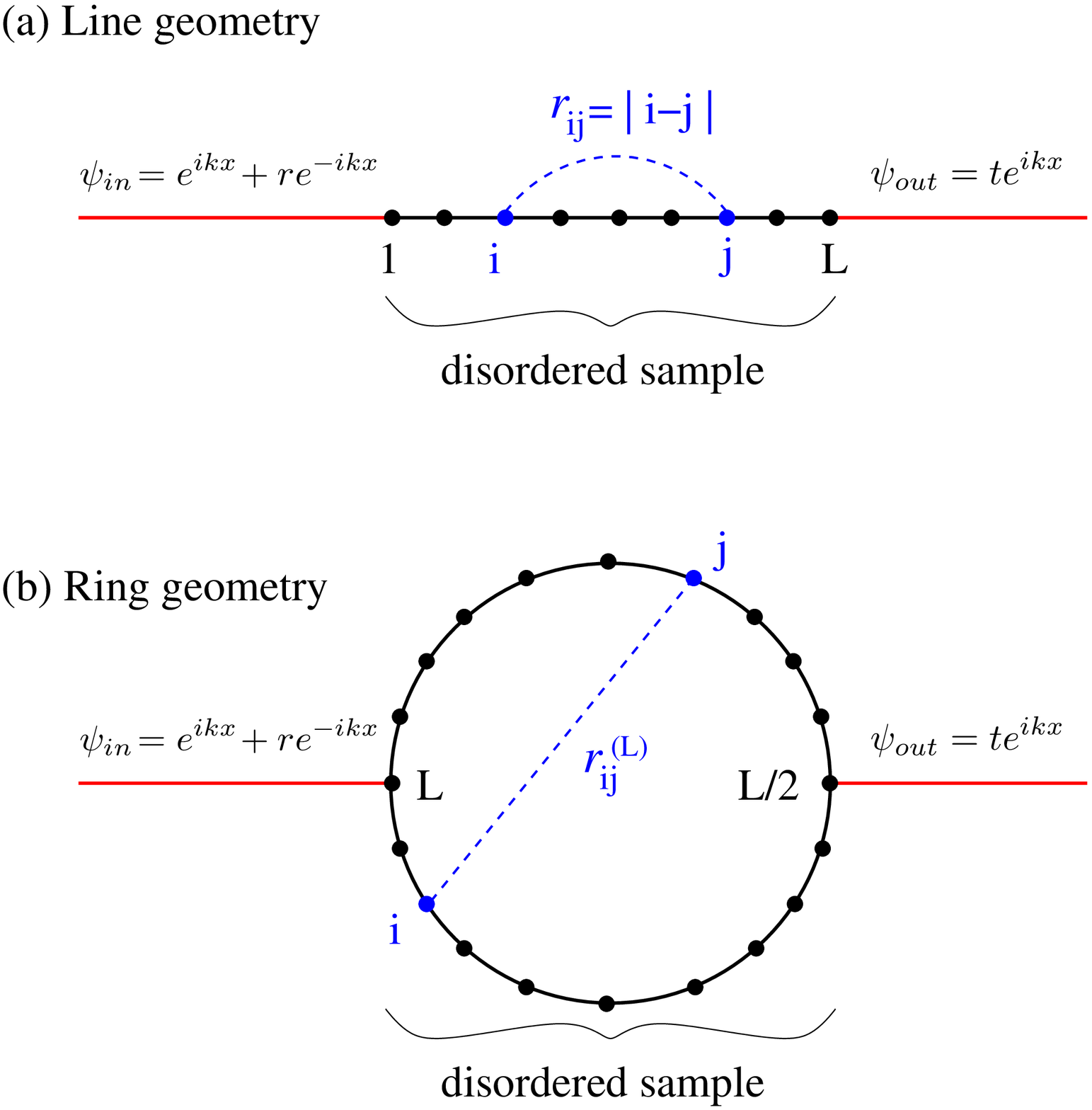}
\caption{ Scattering geometries defining the one-channel transmission 
$T_L^{(1,1)}$ 
(a) Line geometry : the incoming wire is attached to site $1$,
and the outgoing wire to site $L$. These two sites are
'boundary sites' for the disordered sample (they have less neighbors
than bulk sites).
(b) Ring geometry : the incoming wire is attached to site $L$,
and the outgoing wire to site $L/2$. In the ring geometry, all sites 
are 'bulk sites' for the disordered sample (all sites have the same number of neighbors). 
  }
\label{figlinering}
\end{figure}

The one-channel transmission $T_L^{(1,1)}$ 
is the Landauer transmission when the disordered sample
is connected to one incoming wire and one outgoing wire,
see Fig. \ref{figlinering} for the case of the PRBM model.
This  one-channel transmission $T_L^{(1,1)}$
is a very interesting observable 
to characterize Anderson transitions
 \cite{janssen99,evers08} :
(i) it remains finite in the thermodynamic limit 
only in the delocalized phase,
so that it represents an appropriate 
order parameter for the conducting/non-conducting transition;
(ii) exactly at criticality,
 it displays multifractal properties 
in direct correspondence
 with the multifractality of critical eigenstates,
i.e. it displays strong fluctuations
 that are not captured by more global
definitions of conductance.

\subsection{ Multifractal statistics of $T_L^{(1,1)}$ at criticality}

As first discussed in \cite{janssen99}
for the special case of the
 two dimensional quantum Hall transition,
the critical probability distribution of
 the one-channel transmission $T_L^{(1,1)}$
is described by a multifractal spectrum $\Phi^{(1,1)}(\kappa)$
\begin{equation}
{\rm Prob}\left( T_L^{(1,1)} \sim L^{-\kappa}  \right) dT
\oppropto_{L \to \infty} L^{\Phi^{(1,1)}(\kappa) } d\kappa
\label{phikappa}
\end{equation}
Its moments involve non-trivial exponents $X^{(1,1)}(q)$
\begin{equation}
\overline{\left(T_L^{(1,1)}\right)^q}
 \sim \int d\kappa L^{\Phi^{(1,1)}(\kappa) -q \kappa }
\oppropto_{L \to \infty} L^{-X^{(1,1)}(q)}
\label{defXq}
\end{equation}
that can be computed via the saddle-point in $\kappa$
\begin{equation}
-X^{(1,1)}(q) = \opmax_{\kappa} \left[ \Phi^{(1,1)}(\kappa) -q \kappa \right]
\label{saddleXk}
\end{equation}
As stressed in \cite{janssen99}, the physical bound
 $T_L^{(1,1)} \leq 1$
on the transmission implies that the multifractal spectrum
exists only for $\kappa \geq 0$, 
and this termination at $\kappa=0$
 leads to a complete freezing of the moments exponents
 \begin{eqnarray}
X^{(1,1)}(q)  =X^{(1,1)}(q_{sat}) \ \ \ \ {\rm for } \ \ q \geq q_{sat}
\label{freezing}
\end{eqnarray}
at the value $q_{sat}$ where the saddle-point of the integral
of Eq. \ref{defXq} vanishes $\kappa(q \geq q_{sat})=0$.

It is very natural to expect that $\Phi^{(1,1)}(\kappa)$
is related to the multifractal spectrum 
 $f(\alpha)$ concerning eigenfunctions
(see the Appendix \ref{reminder} for a brief reminder).
The possibility proposed in \cite{janssen99} is that before 
the freezing of Eq. \ref{freezing} occurs, 
the transmission should scale as 
the product of two independent
 local densities of states (Eq. \ref{rhomoments})
 \begin{eqnarray}
X^{(1,1)}(q)  = 2 \Delta(q)    \ \ \ {\rm for } \ \ q \leq q_{sat} 
\label{Xqsat}
\end{eqnarray}
We refer to \cite{janssen99} for physical 
arguments in favor of this relation. 
Equations \ref{freezing} and \ref{Xqsat}
 for the moments exponents
are equivalent to following relation between the two
multifractal spectra
\begin{eqnarray}
\Phi^{(1,1)}(\kappa \geq 0) =
 2 \left[ f( \alpha= d+ \frac{\kappa}{2}   ) -d \right]
\label{resphikappa}
\end{eqnarray}
In particular,
the typical transmission
\begin{eqnarray}
   T^{typ}_L \equiv e^{ \overline{ \ln T_L } } 
\label{defTtyp}
\end{eqnarray}
is expected to decay at criticality with some power-law
\begin{eqnarray}
   T^{typ}_L  
\oppropto_{L \to \infty} \frac{1}{L^{\kappa^{(1,1)}_{typ}}}
\label{Ttypcriti}
\end{eqnarray}
where the exponent $\kappa_{typ}^{(1,1)}$ is the point where
 $\Phi^{(1,1)}(\kappa)$
reaches its maximum $\Phi^{(1,1)}(\kappa_{typ}^{(1,1)})=0$
From Eq. \ref{resphikappa}, it is directly related via
\begin{eqnarray}
\kappa_{typ}^{(1,1)} = 2 (\alpha_{typ}- d)
\label{relationxtypalphatyp}
\end{eqnarray}
to the typical exponent $\alpha_{typ}$ 
that characterizes the typical weight of eigenfunctions
\begin{eqnarray}
\vert \psi ( \vec r) \vert^2_{typ} \propto \frac{1}{L^{\alpha_{typ}}}
\label{defalphatyp}
\end{eqnarray}

In our recent work \cite{us_twopoints},
we have tested in detail these predictions
for the statistics of the one-channel transmission $T_L^{(1,1)}$
for the Power-law Random Banded Matrix (PRBM) model where $d=1$
{\it in the ring geometry} (see the scattering geometry 
(b) of Figure \ref{figlinering}),
where all sites are 'bulk sites'.
We thus refer the reader to \cite{us_twopoints}
 for detailed numerical data.

\begin{figure}[htbp]
 \includegraphics[height=6cm]{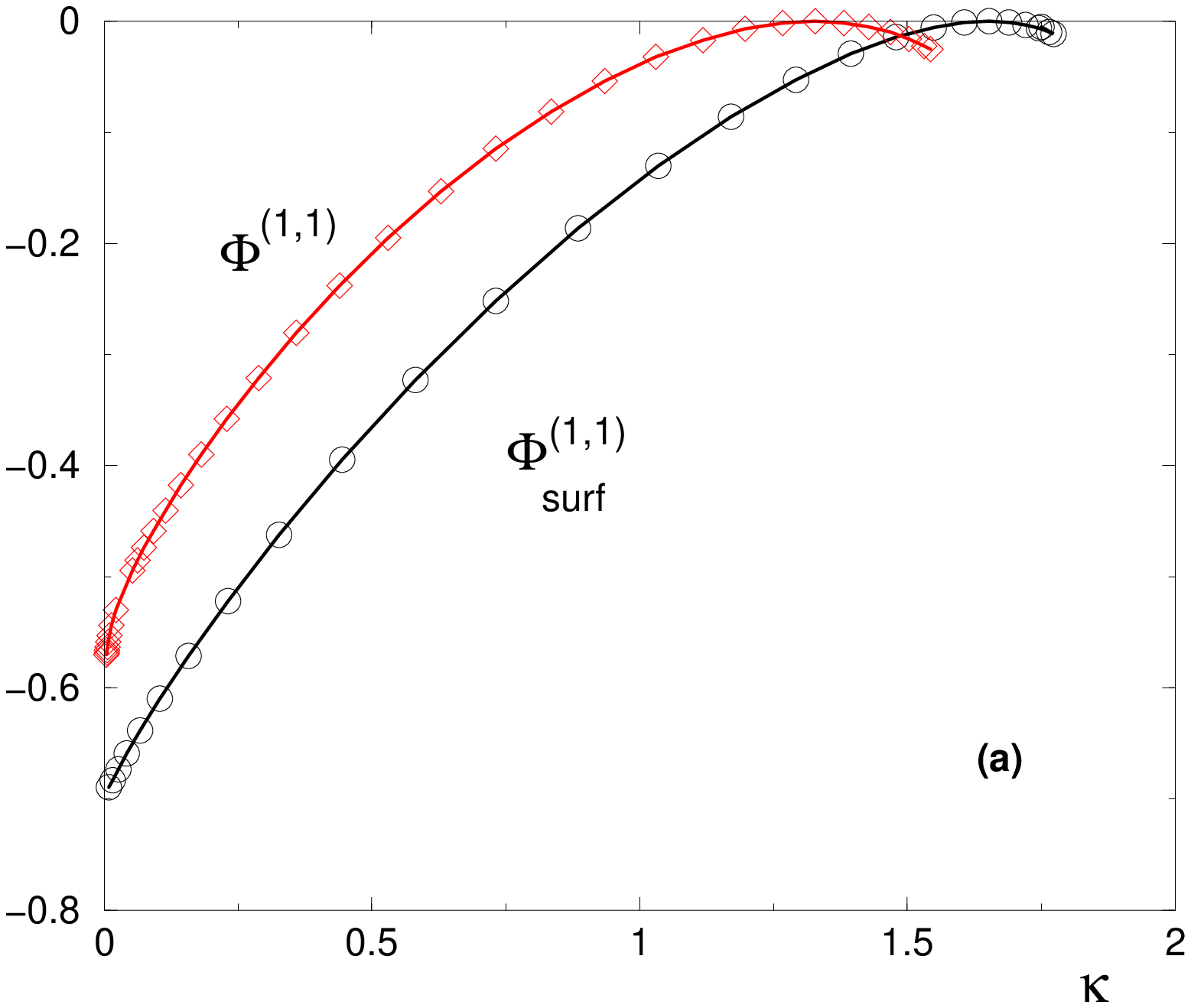}
\hspace{2cm}
 \includegraphics[height=6cm]{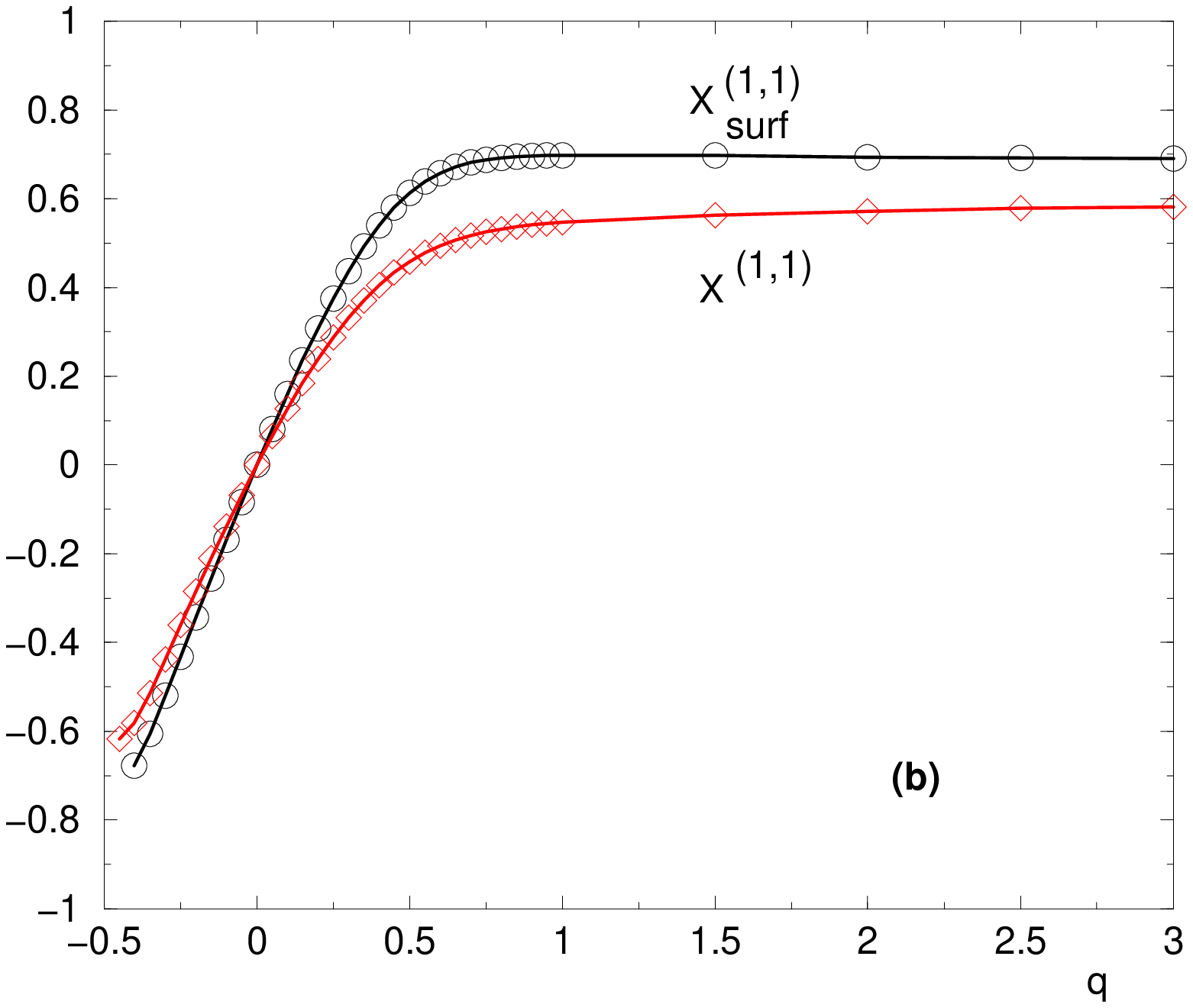}
\vspace{1cm}
\caption{ 
Multifractal statistics of the one-channel transmission $T_L^{(1,1)}$
for $b=0.1$ at criticality $a=1$ 
for the line geometry and for the ring geometry
 (see Fig. \ref{figlinering})
(a) the multifractal spectra $\Phi^{(1,1)}(\kappa)$ 
corresponding to the Ring geometry 
and $\Phi^{(1,1)}_{surf}(\kappa)$ corresponding to the Line
 geometry
are maximal at the typical values $\kappa^{typ} \simeq 1.33$
and $\kappa_{surf}^{typ} \simeq 1.64$.
(b) the corresponding moments exponents $X^{(1,1)}(q)$
and $X^{(1,1)}_{surf}(q)$ saturate at the values 
$X^{(1,1)}(q \geq q_{sat})=-\Phi^{(1,1)}(0) \simeq 0.58$ 
and $X^{(1,1)}_{surf}(q \geq q_{sat})=-\Phi^{(1,1)}_{surf}(0) \simeq 0.69$ }
\label{figsurface}
\end{figure}

However, in other localization models it is usual to attach leads
to the boundaries of the disordered samples.
It is thus interesting to
consider also the scattering geometry (a) of Figure \ref{figlinering}
where the incoming wire
and outgoing wire are attached at boundary points,
since one expects that at criticality, 
boundary points are characterized
by a different multifractal spectrum 
$f_{surf}(\alpha)$ \cite{surf06,surfPRBM,surfhall}.
One then expects that the relation of Eq. \ref{resphikappa}
becomes 
\begin{eqnarray}
\Phi^{(1,1)}_{surf}(\kappa \geq 0) =
 2 \left[ f_{surf}( \alpha= d+ \frac{\kappa}{2}   ) -d_s \right]
\label{resphikappasurf}
\end{eqnarray}
when the transition is measured between two boundary points,
where $d$ is the bulk dimension and $d_s$ is the surface dimension.
Here the line geometry (a) of Figure \ref{figlinering}
corresponds to $d=1$ and $d_s=0$.
On Fig. \ref{figsurface}, we compare 
for the case $b=0.1$ at criticality $a=1$
the bulk multifractal spectrum $\Phi^{(1,1)}(\kappa)$ 
and the surface 
multifractal spectrum $\Phi^{(1,1)}_{surf}(\kappa)$.
We find in particular that the typical exponent 
for the surface case $\kappa^{(1,1)surf}_{typ}(b=0.1) \simeq 1.64$
coincides numerically with the typical value 
of the bulk case $\kappa^{(1,1)}_{typ}(b=0.05) \simeq 1.64$
measured in \cite{us_twopoints}.
This is in agreement with Eq. 29
of Ref. \cite{surfPRBM} stating that in the regime of small $b$,
the surface multifractal spectrum of parameter $b$
coincides with the bulk multifractal spectrum of parameter $b/2$
(here we consider the case $p=0$ in the notations of Ref. \cite{surfPRBM}).

\section{Statistics of the one-to-many-channel transmission
$T_L^{(1,m)}$   }

\label{1tomany}

\subsection{ Scattering geometry }

\begin{figure}[htbp]
 \includegraphics[height=10cm]{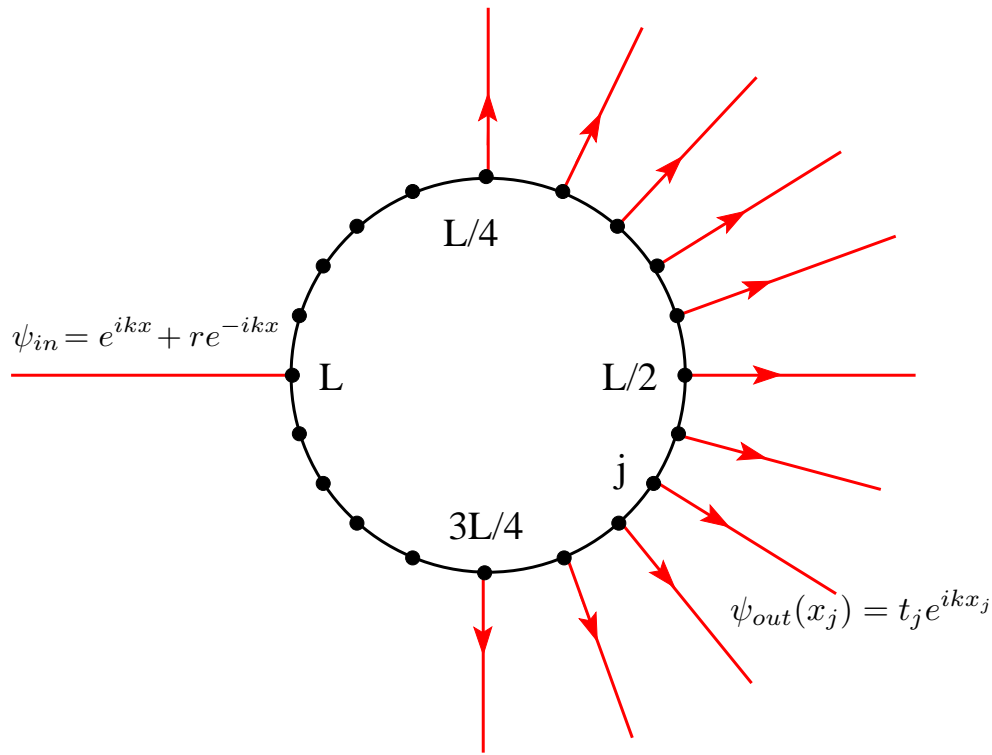}
\caption{ Scattering geometry defining the 'one-to-many' transmission 
$T_L^{(1,m)}$ for the Ring geometry : 
the incoming wire is attached to site $L$,
and outgoing wires are attached at all sites $j$ satisfying
$L/4 \leq j \leq 3L/4$. 
The total transmission is given by Eq. \ref{trans1tomany}
in terms of the amplitudes $t_j$ of the outgoing wires and of the
reflexion amplitude $r$ of the incoming wire.
  }
\label{fig1tomany}
\end{figure}

We now consider the scattering geometry of Fig. \ref{fig1tomany}
where the incoming wire is attached to site $i=L$,
and where there are $(L/2+1)$ outgoing wires attached
to the sites  $j=L/4,L/4+1,...,3L/4$.
We are interested in the transmission
\begin{eqnarray}
T_L^{(1,m)}= \sum_{j=L/4}^{3L/4}  \vert t_j \vert^2 = 1- \vert r \vert^2
\label{trans1tomany}
\end{eqnarray}
where the $t_j$ are the transmission amplitudes
and where $r$ is the reflexion amplitude of the incoming wire
(Eqs \ref{psiwires}).

\subsection{ Multifractal statistics of $T_L^{(1,m)}$ at criticality}

\begin{figure}[htbp]
 \includegraphics[height=6cm]{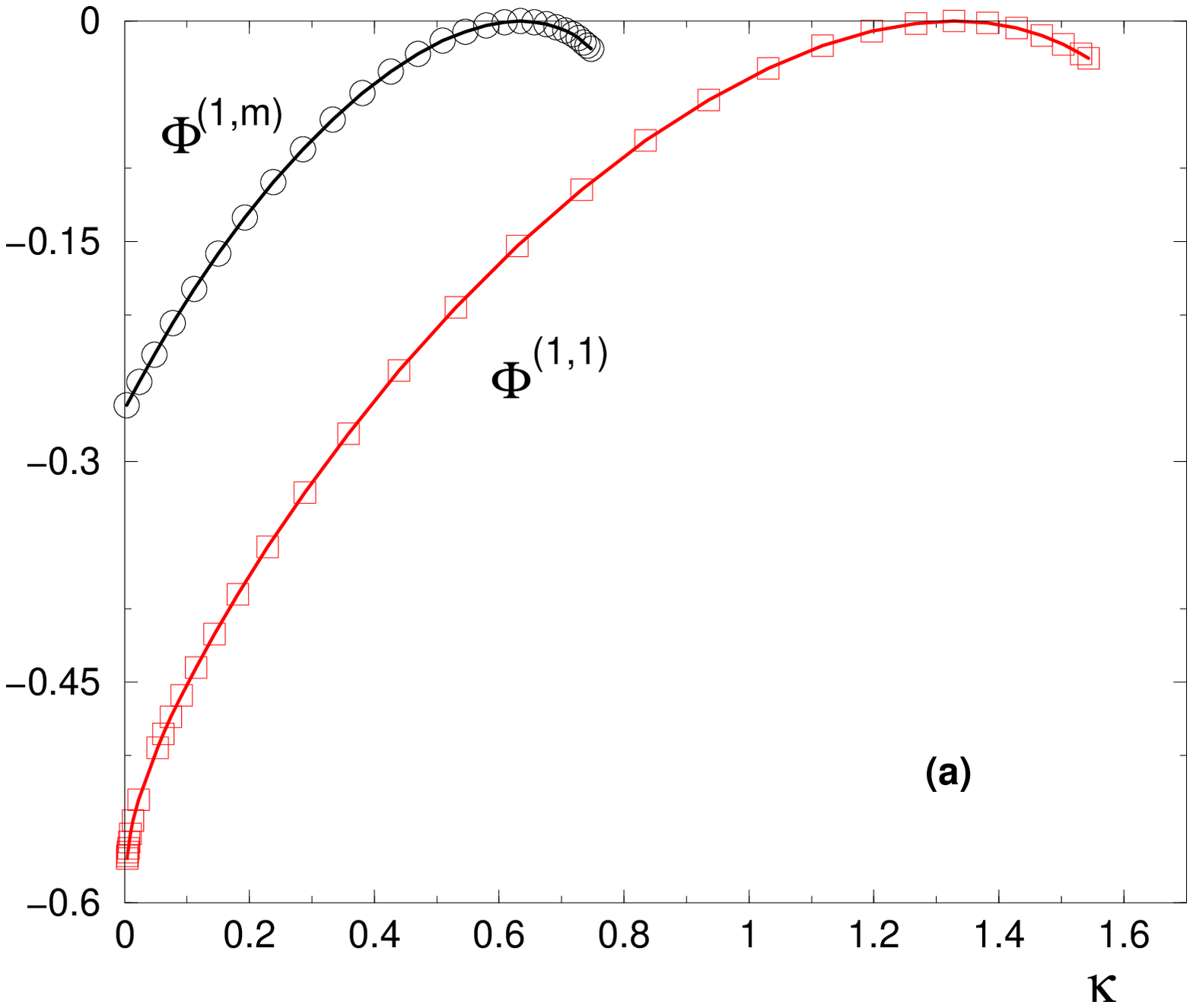}
\hspace{2cm}
 \includegraphics[height=6cm]{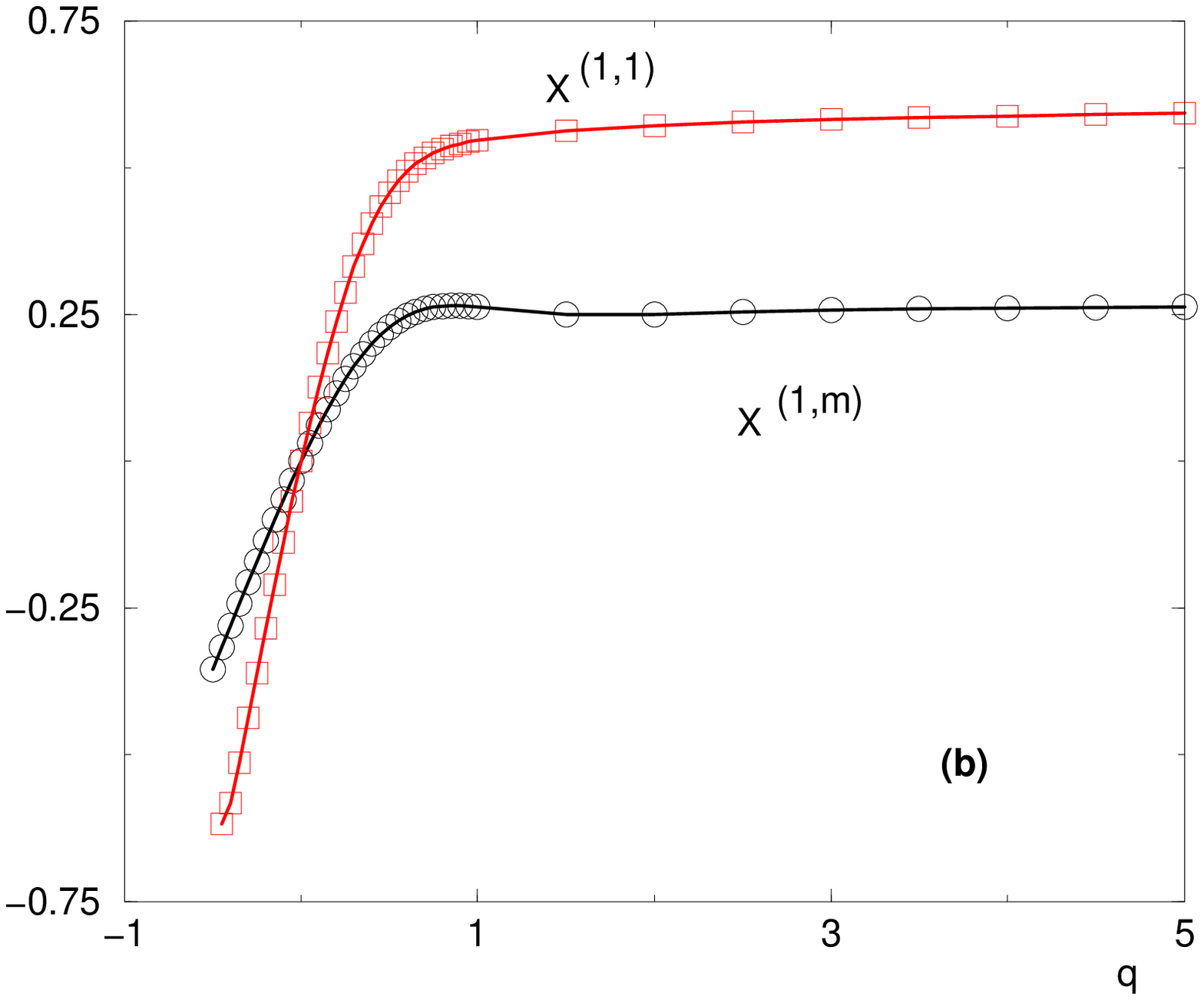}
\vspace{1cm}
\caption{ 
Multifractal statistics of the one-to-many transmission 
for $b=0.1$ at criticality $a=1$ 
for the scattering geometry of Fig. \ref{fig1tomany}
(a) the multifractal spectrum $\Phi^{(1,m)}(\kappa)$
as compared to the multifractal spectrum $\Phi^{(1,1)}(\kappa)$.
The typical values $\kappa_{typ}^{(1,m)} \sim 0.66$ and
$\kappa_{typ}^{(1,1)} \sim 1.32$ satisfy the relation of Eq.
 \ref{typ1m11}. The termination values $\Phi^{(1,m)}(0)\sim -0.27$
and $\Phi^{(1,1)}(0)\sim -0.58$ satisfy the relation of Eq. 
\ref{termin1m11}. The full relation between the 
two multifractal spectra is given by Eq. \ref{resphikappa1m11}.
(b) the corresponding moments exponents $X^{(1,m)}(q)$ 
as compared to $X^{(1,1)}(q )$. }
\label{figmultif1tomany}
\end{figure}

In this scattering geometry, one expects that
 that the critical statistics is still multifractal with a spectrum $\Phi^{(1,m)}(\kappa)$
\begin{equation}
{\rm Prob}\left( T_L^{(1,m)} \sim L^{-\kappa}  \right) dT
\oppropto_{L \to \infty} L^{\Phi^{(1,m)}(\kappa) } d\kappa
\label{phikappa1m}
\end{equation}
and non-trivial exponents $X^{(1,m)}(q)$
\begin{equation}
\overline{\left(T_L^{(1,m)}\right)^q}
 \sim \int d\kappa L^{\Phi^{(1,m)}(\kappa) -q \kappa }
\oppropto_{L \to \infty} L^{-X^{(1,m)}(q)}
\label{defXq1m}
\end{equation}
Again the physical bound
 $T_L^{(1,m)} \leq 1$
on the transmission implies that the multifractal spectrum
exists only for $\kappa \geq 0$, 
and this termination at $\kappa=0$
 leads to a complete freezing of the moments exponents
 \begin{eqnarray}
X^{(1,m)}(q)  =X^{(1,m)}(q_{sat}) \ \ \ \ {\rm for } \ \ q \geq q_{sat}
\label{freezing1m}
\end{eqnarray}
at the value $q_{sat}$ where the saddle-point of the integral
of Eq. \ref{defXq1m} vanishes $\kappa(q \geq q_{sat})=0$.
However in the presence of many wires
(more precisely, whenever the number of outgoing wires is greater than 
$L^{\Phi^{(1,1)}(\kappa=0)/2}$ according
 to the discussion of the previous section), one expects that the 
relation of Eq. \ref{Xqsat} becomes
 \begin{eqnarray}
X^{(1,m)}(q)  =  \Delta(q)    \ \ \ {\rm for } \ \ q \leq q_{sat} 
\label{Xqsat1m}
\end{eqnarray}
meaning that the transmission simply scales as the local density
of states seen by the incoming wire.
Equations \ref{freezing1m} and \ref{Xqsat1m}
 for the moments exponents
are equivalent to following relation 
with the eigenfunction singularity spectrum $f(\alpha)$
\begin{eqnarray}
\Phi^{(1,m)}(\kappa \geq 0) =f( \alpha= d+ \kappa   ) -d 
\label{resphikappa1m}
\end{eqnarray}
Equivalently using Eq. \ref{resphikappa}, one obtain the simple relation
\begin{eqnarray}
\Phi^{(1,m)}(\kappa \geq 0) = \frac{ \Phi^{(1,1)}(2 \kappa)}{2}
\label{resphikappa1m11}
\end{eqnarray}
In particular, the typical values of $\kappa$ (where
the $\Phi$ reach their maximal value 0)
are related by
\begin{eqnarray}
\kappa^{(1,m)}_{typ} = \frac{\kappa^{(1,1)}_{typ}}{2}
\label{typ1m11}
\end{eqnarray}
The termination values at $\kappa=0$ 
are also related by the simple relation
\begin{eqnarray}
\Phi^{(1,m)}( 0) = \frac{ \Phi^{(1,1)}(0)}{2}
\label{termin1m11}
\end{eqnarray}
We have checked these relations (Eqs \ref{resphikappa1m11},
\ref{typ1m11}, \ref{termin1m11}) for the case $b=0.1$
at criticality $a=1$ (see Fig. \ref{figmultif1tomany}).

\section{Statistics of the one-to-many-channel transmission 
$T_L^{(1B,m)}$ in the presence of backward channels }

\label{backwards}

\subsection{ Scattering geometry} 

\begin{figure}[htbp]
 \includegraphics[height=10cm]{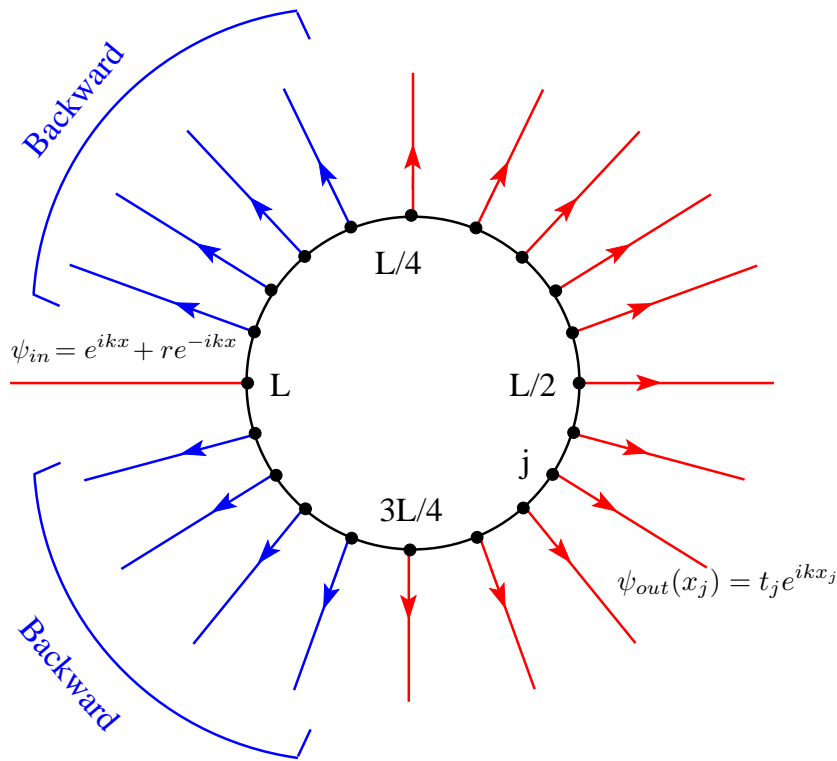}
\caption{ Scattering geometry defining the 'one-to-many' 
transmission 
$T_L^{(1B,m)}$ in the presence of backward channels : 
in addition to Fig. \ref{fig1tomany}, there are now
outgoing wires for $1 \leq j < L/4$ and $3L/4 < j \leq L-1$
that are considered as backward scattering channels.
The total transmission is given by Eq. \ref{transbackwards}
in terms of the amplitudes $t_j$ of the outgoing wires
for $L/4 \leq j \leq 3L/4$.  }
\label{figbackwards}
\end{figure}

We now consider the scattering geometry of Fig. \ref{figbackwards} :
the only difference with the previous case of Fig. \ref{fig1tomany}
is the presence of backward scattering channels near the incoming wire.
We are interested into the transmission
\begin{eqnarray}
T_L^{(1B,m)}= \sum_{j=L/4}^{3L/4}  \vert t_j \vert^2
\label{transbackwards}
\end{eqnarray}
where the $t_j$ are the transmission amplitudes
(Eqs \ref{psiwires}).

\subsection{ Non-multifractal statistics at criticality} 

\begin{figure}[htbp]
 \includegraphics[height=6cm]{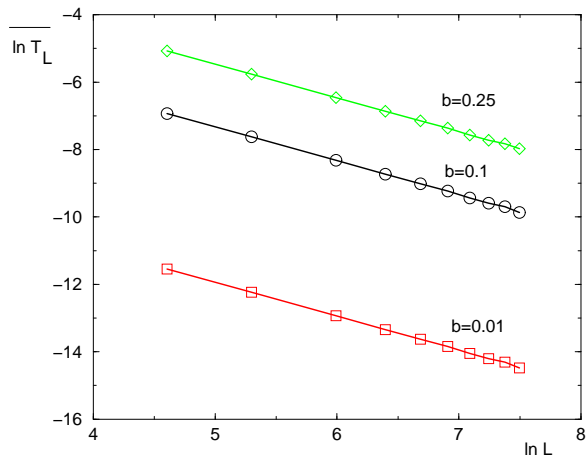}
\vspace{1cm}
\caption{ Statistics of the transmission $T_L^{(1B,m)}$
for the scattering geometry of Fig. \ref{figbackwards}
  at criticality $a=1$ :
 $ \ln T_L^{typ}\equiv  \overline{ \ln T_L } $
as a function of $ \ln L$.
The slopes  for the three values $b=0.01$, $b=0.1$
and $b=0.25$ coincide $\kappa_{back}(b)=1 $
(see Eq. \ref{kappamono}). }
\label{figtypback}
\end{figure}

In this case, we find numerically that the statistics of the 
transmission is not multifractal anymore, but mono-fractal
\begin{eqnarray}
T_L^{(1B,m)} \equiv \frac{\tau}{L^{\kappa_{back}}}
\label{monbackwards}
\end{eqnarray}
where $ \tau$ is a random variable of order $O(1)$.
Moreover, the monofractal exponent is
\begin{eqnarray}
\kappa_{back} (b) = 1  \ \ {\rm for \ \ all } \ \ b
\label{kappamono}
\end{eqnarray}
as shown on Fig. \ref{figtypback} for the three values 
$b=0.01$, $b=0.1$ and $b=0.25$, i.e.
it does not depend anymore on the multifractal critical properties
that are known to
change continuously with the parameter $b$ in the PRBM.
For instance for the one-channel transmission, we have measured
in \cite{us_twopoints} the typical exponents
$\kappa^{(1,1)}_{typ}(b=0.01) =1.92 $,
$\kappa^{(1,1)}_{typ}(b=0.1) =1.33 $, and
$\kappa^{(1,1)}_{typ}(b=0.25) = 0.77$ which are significantly different.

\subsection{ Consequences for the 
usual many-channel (m.c.) transmission $T_L^{m.c.}$ }

\label{manychannel}

For the short-range Anderson tight-binding model in dimension $d=3$,
the usual many-channel 
Landauer transmission $T_L^{m.c.}$ 
corresponds to the incoherent
sum 
\begin{eqnarray}
T_L^{m.c.} = \sum_{i \in L^{d_s}} T_i
\label{tmc}
\end{eqnarray}
of $L^{d_s}=L^{d-1}$ contributions $T_i$,
where $T_i$ is the coherent scattering transmission 
for one-incoming wire arriving at $i$,
with $L^{d_s}$ backward scattering channels 
and with $L^{d_s}$ forward scattering channels
(see for instance Fig. 2 of Ref \cite{buttiker}).
The statistics of this many-channel transmission
has been much studied (see the review \cite{markos}) :
$T_L^{m.c.}$ is a random variable of order $O(1)$.
One of our motivations for the present study was
to determine whether the statistics of each contribution $T_i$
was multifractal or not.
Since each contribution $T_i$ is analog to the transmission of
scattering geometry of Fig. \ref{figbackwards},
our present study suggests that 
each contribution $T_i$ in the incoherent sum of Eq. \ref{tmc}
is not multifractal, but monofractal with an exponent 
$\kappa_{back}=d_s=d-1$ in short-range models,
i.e. this exponent does not depend on the critical
multifractal spectrum.

\section{ Influence of the scattering geometry on the transmission off criticality }

\label{seclocdeloc}

Up to now, we have only discussed the transmission statistics
at criticality. For completeness, we briefly describe in this section
how the behavior of the transmission depends on the scattering geometry
outside criticality.

\subsection{ Influence of the scattering geometry in the localized phase  }

\begin{figure}[htbp]
 \includegraphics[height=6cm]{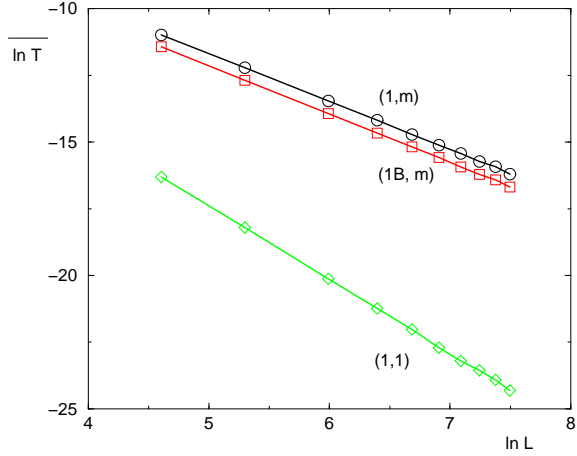}
\caption{ Statistics of the transmissions 
$T^{(1,1)}_L$, $T^{(1,m)}_L$ and $T^{(1B,m)}_L$ 
for the scattering geometries of Fig. \ref{figlinering}, \ref{fig1tomany}
and \ref{figbackwards}
 in the localized phase $a=1.4$ for $b=0.1$ :
the measured slopes on this log-log plot are in agreement
with Eq. \ref{tloc}. }
\label{figloc}
\end{figure}

In usual short-range models, the localized phase
is characterized by exponentially localized wavefunctions,
whereas in the presence of power-law hoppings, 
localized wavefunction can only decay with power-law integrable tails.
For the PRBM, it is moreover expected that the asymptotic
decay is actually given exactly by the power-law of 
Eq. \ref{defab} for the hopping term defining the model \cite{mirlin96} :
\begin{eqnarray}
  \vert \psi (r) \vert^2_{typ} \sim\frac{1}{r^{2a} }
\label{psiloc}
\end{eqnarray}
As a consequence in the localized phase $a>1$, one expects
 the following typical decays for the transmissions
$T^{(1,1)}_L$, $T^{(1,m)}_L$ and $T^{(1B,m)}_L$ 
for the scattering geometries of Fig. \ref{figlinering}, \ref{fig1tomany}
and \ref{figbackwards}
\begin{eqnarray}
   \overline{ \ln T^{(1,1)}_L(a>1) }&& \oppropto_{L \to \infty}  - 2 a \ln L 
\nonumber \\
  \overline{\ln T^{(1,m)}_L(a>1) }&& \oppropto_{L \to \infty}  - (2 a-1) \ln L 
\nonumber \\
  \overline{\ln  T^{(1B,m)}_L(a>1)} && \oppropto_{L \to \infty}
  - (2 a-1) \ln L
\label{tloc}
\end{eqnarray}
The decay of $T^{(1,1)}_L$ directly reflects the decay of Eq. \ref{psiloc}
of wavefunctions. The presence of many outgoing channels
change the power by one as a consequence of the integration
over the all the sites connected to outgoing channels.
And the presence of backward channels does not change this exponent.
This is in agreement with our numerical results shown on Fig. 
\ref{figloc} for the case $a=1.4$ and $b=0.1$.
We refer to our previous work \cite{us_twopoints} for 
more details on the histogram around the typical value
for the one-channel case.

\subsection{ Influence of the scattering geometry in the delocalized phase  }

In the delocalized phase, the eigenfunctions 
are not multifractal, but monofractal with the 
single value $\alpha_{deloc}=d$ for the weight $\vert\psi_L^2(x)\vert$.
As a consequence, the typical transmission for the one-channel case
of Fig. \ref{figlinering}
is expected to remain finite as $L \to +\infty$ \cite{janssen99}
(in Eq. \ref{relationxtypalphatyp}, the case $\alpha_{typ}=d$ yields 
$\kappa_{typ} = 0$)
\begin{eqnarray}
   T^{(1,1)}_L (a<1)
\oppropto_{L \to \infty} T^{(1,1)}_{\infty} >0
\label{T11deloc}
\end{eqnarray}
The one-channel transmission is thus a good order parameter
of the transport properties \cite{janssen99}.
We refer to our previous work \cite{us_twopoints} for 
more details on the histogram 
for this one-channel case.
In the presence of many out going wires as in Fig. \ref{fig1tomany},
the transmission will thus also remains finite
\begin{eqnarray}
   T^{(1,m)}_L (a<1)
\oppropto_{L \to \infty} T^{(1,m)}_{\infty} >0
\label{T1mdeloc}
\end{eqnarray}
However in the presence of backward scattering channels
as in Fig. \ref{figbackwards}, one expects that the transmission 
will now decay with $L$.
For instance in the usual short-range Anderson model in dimension $d=3$,
the many-channel transmission of Eq. \ref{tmc} scales as $L^{d-2}$,
i.e. each contribution $T_i$ in the presence of backward channels
scales as $1/L$, which represents the probability to reach first
the out-going surface without returning to the incoming surface
for a diffusive particle.
For the PRBM, our numerical results indicate the typical decay
\begin{eqnarray}
 \overline{\ln  T^{(1B,m)}_L(a<1)} && \oppropto_{L \to \infty}
  - (2 a-1) \ln L  
\label{T1Bmdeloc}
\end{eqnarray}
which corresponds to the probability to make a direct hopping
towards a forward channel without visiting the sites connected
to the backward channels.

\section{Conclusion}

\label{secconclusion}

In this paper, we have studied how the scattering geometry 
determines the statistics of the Landauer transmission,
both at criticality and outside criticality.
We have presented detailed numerical results for the PRBM model
for various scattering geometries that interpolate
between the two cases that are usually considered,
namely the one-channel and the many-channel cases.
We have found that :
(i) in the presence of one isolated incoming wire
and many outgoing wires, the transmission 
has the same multifractal statistics as the local density of states
of the site where the incoming wire arrives;
(ii) in the presence of backward scattering channels
 with respect to the case (i), the statistics 
of the transmission is not multifractal anymore, but becomes monofractal.

\appendix

\section{ Reminder on multifractality of eigenfunctions}

\label{reminder}

In this Appendix, we recall some useful properties
concerning eigenfunction multifractality.
 The multifractal spectrum $f(\alpha)$ is defined as follows
(for more details see for instance the reviews 
\cite{janssenrevue,mirlinrevue}):
in a sample of size $L^d$, the number ${\cal N}_L(\alpha)$
of points $\vec r$ where the weight $\vert \psi(\vec r)\vert^2$
scales as $L^{-\alpha}$ behaves as 
\begin{eqnarray}
{\cal N}_L(\alpha) \oppropto_{L \to \infty} L^{f(\alpha)}
\label{nlalpha}
\end{eqnarray}
The typical value $\alpha_{typ}$
 corresponds to the maximum value $f(\alpha_{typ})=d$
of the multifractal spectrum $f(\alpha)$.
The inverse participation ratios (I.P.R.s) can be then rewritten
as an integral over $\alpha$
\begin{equation}
Y_q(L)  \equiv \int_{L^d} d^d { \vec r}  \vert \psi (\vec r) \vert^{2q}
\simeq \int d\alpha \ L^{f(\alpha)} 
\ L^{- q \alpha} \opsimeq_{L \to \infty} 
L^{ - \tau(q) }
\label{ipr}
\end{equation}
The exponent $\tau(q)$ can be obtained via a saddle-point
calculation in $\alpha$, and one obtains the Legendre
transform formula \cite{janssenrevue,mirlinrevue}
\begin{eqnarray}
 q && =f'(\alpha) \nonumber \\
 \tau(q) && =  q \alpha  - f(\alpha)
\label{legendre}
\end{eqnarray}
These scaling behaviors, which concern  
individual eigenstates $\psi$, can be translated
for the local density of states
\begin{equation}
\rho_L(E,\vec r) = \sum_{n} \delta(E-E_n) \vert \psi_{E_n}(\vec r)\vert^2
\label{defrho}
\end{equation}
as follows : for large $L$, 
when the $L^d$ energy levels become dense,
the sum of Eq. \ref{defrho} scales as
\begin{equation}
\rho_L(E, \vec r) \propto L^d \vert \psi_E(\vec r)\vert^2
\label{equiv}
\end{equation}
and its moments involve the exponents
 $\tau(q)$ introduced in Eq. \ref{ipr}
\begin{equation}
\overline{ [\rho_L(E,\vec r)]^q } \oppropto_{L \to \infty}
\frac{1}{L^{\Delta(q)}} \ \ {\rm with \ \ } \Delta(q) =  \tau(q)-d (q-1) 
\label{rhomoments}
\end{equation}

These notions concerning bulk multifractality have been
recently extended to surface multifractality 
\cite{surf06,surfPRBM,surfhall} : the idea is that 
points near the boundaries are described by another
multifractal spectrum $f_{surf}(\alpha)$
which is in general not simply related to the bulk
multifractal spectrum $f(\alpha)$.

\end{document}